\documentclass[reprint,aps,floatfix,superscriptaddress]{revtex4-2}
\usepackage{amsmath,amssymb,amsthm}
\usepackage{graphicx}
\graphicspath{{figures/}{./}}
\usepackage{booktabs}
\usepackage{longtable}
\usepackage{xcolor}
\usepackage[hidelinks]{hyperref}
\usepackage{rotating}
\counterwithout*{equation}{section}

\newtheoremstyle{thmnoparen}{6pt}{6pt}{\itshape}{}{\bfseries}{.}{ }{\thmname{#1}\thmnumber{ #2}\thmnote{ #3}}
\theoremstyle{thmnoparen}

\newtheorem{remark}{Remark}

\newcommand{\Pp}{P_{p}}

\begin{document}
\title{The wiring sets a phase-blind quantum memory's gap, the weight caps its coherence}
\author{Ruini Qian}
\affiliation{Faculty of Computility Microelectronics, Shenzhen University of Advanced Technology, Shenzhen, Guangdong 518107, China}
\affiliation{Guangdong Provincial Key Laboratory of Computility Microelectronics, Shenzhen, Guangdong 518107, China}
\author{Zhaobin Lyu}
\affiliation{Pengxin Quantum Technology (Shenzhen) Co., Ltd., Shenzhen 518107, China}
\author{Zelong Yin}
\affiliation{E. L. Ginzton Laboratory and the Department of Applied Physics, Stanford University, Stanford, California 94305, USA}
\author{Jingjing Hu}
\email{hujingjing@suat-sz.edu.cn}
\affiliation{Faculty of Computility Microelectronics, Shenzhen University of Advanced Technology, Shenzhen, Guangdong 518107, China}
\affiliation{Guangdong Provincial Key Laboratory of Computility Microelectronics, Shenzhen, Guangdong 518107, China}
\affiliation{Pengxin Quantum Technology (Shenzhen) Co., Ltd., Shenzhen 518107, China}
\author{Dengfeng Li}
\email{lidengfeng@suat-sz.edu.cn}
\affiliation{Faculty of Computility Microelectronics, Shenzhen University of Advanced Technology, Shenzhen, Guangdong 518107, China}
\affiliation{Guangdong Provincial Key Laboratory of Computility Microelectronics, Shenzhen, Guangdong 518107, China}
\affiliation{Pengxin Quantum Technology (Shenzhen) Co., Ltd., Shenzhen 518107, China}
\author{Shuoming An}
\email{anshuoming@suat-sz.edu.cn}
\affiliation{Faculty of Computility Microelectronics, Shenzhen University of Advanced Technology, Shenzhen, Guangdong 518107, China}
\affiliation{Guangdong Provincial Key Laboratory of Computility Microelectronics, Shenzhen, Guangdong 518107, China}
\affiliation{Pengxin Quantum Technology (Shenzhen) Co., Ltd., Shenzhen 518107, China}
\begin{abstract}Engineered dissipation corrects errors, yet a memory's phase decays. What the memory's gap measures is not settled. Here we show that when each jump resets one basis state, the wiring can set the gap time while the phase keeps its rate. Under dephasing the gap time spans two orders across wirings; the phase does not. In this class, a contrast on two or more qubits fades faster than a bare qubit at any distance; no such pump power substitutes. The ceiling is two over the qubits the contrast spans. A two-branch pump can escape it at distance three; a two-clock test runs on published data.\end{abstract}

\maketitle

\section{Introduction}
\label{sec:intro}
Engineered dissipation holds a code space without measurement or feedback, the protected subspace that errors would drive the encoded state out of. Stabilizer codes, whose checks measure products of Pauli operators, take the measurement route instead~\cite{acharya2023,livingston2022}. Demonstrated in trapped ions, superconducting circuits, and bosonic modes~\cite{barreiro2011,shankar2013,leghtas2015,touzard2018,gertler2021,chen2025sustech,li2024star}, engineered dissipation has reached break-even, an encoded lifetime that exceeds one bare qubit's coherence~\cite{brock2025,ni2025,li2025,shirol2026,sun2025}, as has measurement-based feedback~\cite{ofek2016,sivak2023}.

Two clocks already run in one device: on the driven-dissipative cat qubit, two-photon dissipation holds a bit-flip time above ten seconds~\cite{reglade2024}, saturating where single-mode theory predicts growth~\cite{ferrari2026}, while single-photon loss sets the phase-flip time above $490\;\mathrm{ns}$~\cite{reglade2024}. They are the analogs of a qubit's $T_1$ and $T_2$, and run independently though a physical qubit ties the pair by $T_2 \le 2T_1$.

Such a bath's Liouvillian gap is naturally read as the memory's protection. The gap is the bath generator's slowest nonzero rate, $\min|\mathrm{Re}\,\lambda|$ over its eigenvalues besides the steady state, its inverse the gap time $T_{\rm gap}$; the protection is the stored phase's lifetime $T_X$. A pump, a bath whose every jump resets one computational-basis error, is wiring: each error a leak out of the code space, each jump the wire that carries it back. The layout of the wires, not the pressure behind them, sets how long the code space takes to refill, and no wire carries the phase.

Prior work leaves the gap's magnitude open: a constant gap does not by itself give a growing memory time~\cite{shtanko2025}; relaxation-time figures of merit fail for logical qubits~\cite{pal2022}; and one decay rate cannot describe a memory needing two conjugate observables~\cite{pastawski2011}. Which modes reach a logical channel, decoder included, is characterized~\cite{gordeychuk2026}; the rate the phase is left at is bounded here, beyond the undetectability argument~\cite{knill1997,terhal2015}. Earlier work tunes the gap by a boundary term~\cite{li2026gapcontrol} or places it among neighboring slow modes~\cite{labaymora2026}; here the wiring sets it.

Here we show that when dephasing dominates, the gap's magnitude is the pump's wiring, not its protection, and the phase keeps its unpumped rate. The wiring maps an error's state to the code state it returns to. On a four-qubit plaquette carrying one check, at damping time $T_1 = 200\;\mu$s and dephasing time $T_\phi = 100\;\mu$s, the $8!$ error-to-code wirings put $T_{\rm gap}$ from $200$ to $40300\;\mu\mathrm{s}$ while the coherence holds near $20\;\mu\mathrm{s}$.

\begin{figure}[!htbp]
\centering
\includegraphics[width=\columnwidth]{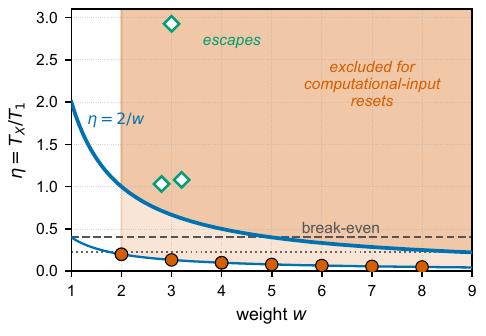}
\caption{The phase-contrast boundary. Horizontal axis, the logical weight $w$, the number of qubits on which the two codewords differ; vertical axis, $\eta\equiv T_X/T_1$, the logical-coherence lifetime in units of the bare damping time $T_1$. For a weight-$w$ coherence sector under resets that take one basis state at a time, independent dephasing and $T_1$ decay, the excluded region, for $w\ge2$, extends above the lower blue curve, $\eta=T_2/(wT_1)=0.4/w$, the tighter of the two bounds; the lighter corridor up to the upper curve is the share the dephasing accounts for. Vermillion markers are the class's contrasts at dephasing time $T_\phi=100\;\mu\mathrm{s}$ and damping time $T_1=200\;\mu\mathrm{s}$: the two-plaquette array and the $w=2$ repetition chain, the plaquette and the $w=4$ chain, and the chains through $w=8$, tracing $\eta=0.4/w$; the dashed line is one bare qubit's coherence under that noise, $\eta=0.4$. Green diamonds are protocols outside these conditions, all at weight three, offset by $\pm0.2$ in $w$ for legibility: one jump returning both error branches, the rank-two phase-preserving calculations of Sec.~S3, at pump rate $\kappa=2$ MHz, $T_\phi=50\;\mu\mathrm{s}$ and $T_1=200\;\mu\mathrm{s}$. Left to right, the Steane code at $\eta=1.03$, the five-qubit code at $2.92$, surface-17 at $1.08$; the dotted line is their own bare-qubit reference, $\eta=0.222$.}\label{fig:boundary}
\end{figure}

The coherence answers to weight instead. A contrast of logical weight $w$, the coherence between two codewords differing on $w$ qubits, fades no slower with the pump than without: its trace distance, how distinguishable the two are, never exceeds its unpumped value. That holds for zero Hamiltonian, independent dephasing on each qubit at rate $\gamma_\phi=1/(2T_\phi)$, damping at time $T_1$, and basis-state resets. Its lifetime $T_X$ obeys
\begin{equation}
\eta \equiv \frac{T_X}{T_1} \le \frac{2}{w+4wT_1\gamma_\phi} \le \frac{2}{w}.
\label{eq:eta_bound}
\end{equation}
The middle expression is the bare qubit's own line, $\eta_{\rm bare} = T_2/T_1$ with $1/T_2 = 2\gamma_\phi + 1/(2T_1)$, divided by the weight: at $w=1$ the ceiling matches one qubit's relation. At $w\ge2$ no contrast of this class outlives one bare qubit at any code distance, the smallest number of qubits on which two codewords differ, Fig.~\ref{fig:boundary}.

\section{What the pump reads}
\label{sec:separation}
\label{sec:phase}
A pump's class is fixed by three things: the sectors the noise reaches, what the input carries, and how its branches return to the code. Let $g$ be a stabilizer generator with code space $\mathcal{C}$, complement $\mathcal{C}^\perp$, and projectors $P_\pm = (I \pm g)/2$, $P_{\mathcal C}=P_+$. In the dissipative state-preparation framework~\cite{kraus2008,diehl2008}, we take the simplified Diehl--Kraus pump of Sec.~S7~\cite{sm}: rank-one jumps $L_\alpha = \sqrt{\kappa}\,|\psi_\alpha\rangle\langle\phi_\alpha|$ at rate $\kappa$, $|\phi_\alpha\rangle \in \mathcal{C}^\perp$, $|\psi_\alpha\rangle \in \mathcal{C}$.

The pump acts only across the code-space boundary: $L_\alpha P_+=0$ implies the Lindblad dissipator $\mathcal D[L_\alpha](P_+\rho P_+)=0$, with $\mathcal{D}[L]\rho=L\rho L^\dagger-\tfrac12\{L^\dagger L,\rho\}$ the loss-and-jump map of a Lindblad channel~\cite{poyatos1996,lindblad1976}, App.~\ref{sec:appF}. For the lifetime bound, each $|\phi_\alpha\rangle$ must be a computational-basis error ket $|e_\alpha\rangle$; the code targets may be superpositions. We call such a pump \emph{phase-blind} with respect to computational coherences. Noise the pump cannot affect inside the code space is \emph{pump-invisible}.

\begin{remark}[Stabilizer-preserving noise is pump-invisible]
\label{prop:main}
Let every Kraus component $E$ of the noise channel satisfy $[E, g] = 0$ on $\mathcal{C}$, equivalently $E\mathcal{C} \subseteq \mathcal{C}$. Then the noise is invisible to the pump $\{L_\alpha\}$: its action within the code space is uncorrected, and the logical coherence decays at the rate the noise sets there, unconstrained by the engineered rate $\kappa$. The noise channel preserves the code-space block: the completeness relation $\sum_kE_k^\dagger E_k=I$ cancels each component's outflow, so inside the code space the dissipative part of the generator is the noise's alone. For continuous evolution the support condition must hold at every time.
\end{remark}

The identity holds for an arbitrary Hermitian projector and is machine-checked in Lean~4, App.~\ref{sec:appF}. Residual dissipators from adiabatic elimination enter as an added in-code rate, Sec.~S10.

Jump rank and code darkness are the obvious candidates for the criterion behind Eq.~\eqref{eq:eta_bound}, and neither suffices. A code is dark when every jump annihilates the code space. Take three qubits with no Hamiltonian and four equal-rate rank-one jumps whose error bras and code targets are complete orthonormal families. The pump annihilates the code space even at the minimal rank, yet superposition-valued error inputs shuttle phase into population and back, so what survives is a classical label rather than a memory, outliving the $w=2$ ceiling by $1.580\times$, Sec.~S6. That is why the class is restricted to computational-basis inputs.

The criterion is the input structure, not the rank alone. A jump whose error bra is a computational-basis state cannot match both basis strings of an off-diagonal contrast, so it cannot create that coherence. The phase survives only when a single jump reaches both strings, as a superposed error input or a coherent return does. A rank-one jump returns one error branch to a fixed code state and cannot create the error-code coherence $|e\rangle\langle c|$, which the damping shortens. Computational-basis targets create diagonal population, superposition targets prepare the pump's own phase rather than the state's, and separate reset jumps erase the error coherence $|e_0\rangle\langle e_1|$. One coherent correction jump returns the code coherence $|c_0\rangle\langle c_1|$, the autonomous quantum-error-correction dissipator $\sqrt\kappa P_{\mathcal C}E^\dagger$~\cite{cohen2014,reiter2017,shtanko2025}.

That asymmetry is why commuting noise damps the coherence at its unpumped rate at any speed, Secs.~S6 and S7. The Kapit shadow lattice~\cite{kapit2015}, an auxiliary lattice whose local couplings return each error to the code, is the canonical example; Remark~\ref{prop:main} elevates it to any stabilizer whose noise commutes with the check. For a $Z$-type code under dephasing the syndrome, the pattern of check outcomes, does not change, so no syndrome-based correction can act on the coherence~\cite{knill1997,terhal2015}. The undetectability is the code's, not the pump's: measurement-based recovery gives the same rate, Sec.~S3.

\section{The wiring sets the gap time, not its protection}
\label{sec:kapit}
The separation between the gap and the coherence is cleanest on a four-qubit plaquette, the elementary cell of the driven-dissipative toric code~\cite{kapit2015}, built on Kitaev's toric code~\cite{kitaev2003}. The toric code's checks act on stars and plaquettes. We take the Hadamard dual of the usual cell, the relabeling $X\leftrightarrow Z$ on every qubit. Its plaquette stabilizer is then $\Pp = Z_1 Z_2 Z_3 Z_4$, the $Z$-type form that a $Z$-parity readout measures natively, Sec.~S10. $\sigma^z$ dephasing commutes with it, while the $X$-type construction puts $\sigma^z$ in the syndrome-flipping sector, where errors change what a check reads, collapsing the separation.

The pump is a Diehl--Kraus surrogate for this class's reset, rank-one jumps obtained by eliminating a pumped ancilla, at $\kappa = 2\;\mathrm{MHz}$; it is not the shadow lattice. Each computational-basis $\Pp = -1$ eigenstate maps to a code state under the cyclic wiring $\mathcal{W}$ that this work adopts, its fixed error-to-code assignment: the $i$-th error state in increasing binary order goes to the $i$-th of the listing $|0000\rangle$, $|1001\rangle$, $|1010\rangle$, \dots, Sec.~S7.

The parameters are $T_1 = 200\;\mu\mathrm{s}$ and pure dephasing at $T_\phi = 100\;\mu\mathrm{s}$, so $\sigma^z$ dominates. At zero Hamiltonian the Liouvillian's zero eigenspace is 64-fold~\cite{albert2014}, so the pump alone sets no timescale; the noise lifts the degeneracy to the unique steady state $|0000\rangle$~\cite{ticozzi2014}. The Hamiltonian is tolerated up to code-error drive amplitudes $h\lesssim\kappa$, $h$ the coefficient of a single-qubit term $hX_1$, Sec.~S6.

\begin{figure}[t]
\centering
\includegraphics[width=\columnwidth]{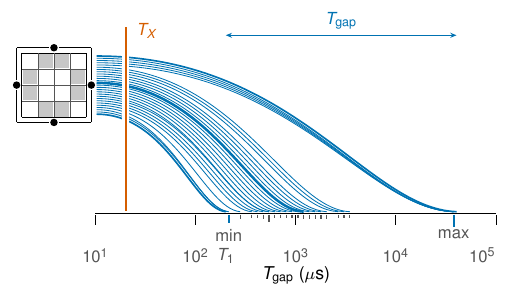}
\caption{The wiring is how the pump matches each error to a code state; it sets the refill time. The noise sets how long the phase lasts. The plaquette at the left carries the check $Z_1Z_2Z_3Z_4$, one qubit on each edge: its sixteen cells are the sixteen states of those four qubits, eight shaded failing the check and eight open passing it, so of the $8!=40320$ ways to match a failing state to a passing one, nearly all refill on a real timescale, the rest returning pairs that the exact computation resolves to one real mode. The ways are drawn as one curve per occupied time bin, the curve of this work's wiring the heaviest. The vertical spread is a rendering, not data. The blue arrow spans the gap times $T_{\rm gap}$ of all $8!$ wirings, and the strokes below the axis count the ways in each bin. The fastest ways sit at the bare damping time $T_1$, and the vermillion rule marks the encoded phase's lifetime $T_X$: the contrast spans four qubits, so it lasts about $20\;\mu$s whatever the wiring.}
\label{fig:census}
\end{figure}

\begin{figure}[t]
\centering
\includegraphics[width=\columnwidth]{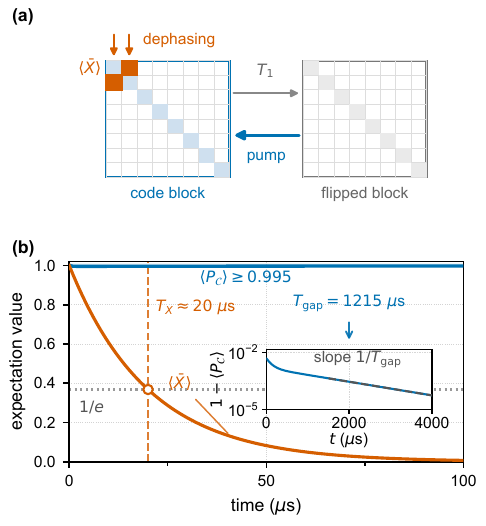}
\caption{How long the code space stays populated is not the logical lifetime on the $Z$-type plaquette, a four-qubit code with stabilizer $Z_1Z_2Z_3Z_4$ and conjugate logical $\bar X=X_1X_2X_3X_4$, at pump rate $\kappa=2\;\mathrm{MHz}$, amplitude-damping time $T_1=200\;\mu\mathrm{s}$ and pure dephasing at $T_\phi=100\;\mu\mathrm{s}$. (a) The density matrix as a schematic, two sectors. Blue is the code block, where the stabilizer reads $+1$, grey its $-1$ counterpart, each block's populations on the diagonal. The pump cycles population between them, setting the gap time $T_{\rm gap}=1215\;\mu\mathrm{s}$ of the cyclic wiring this work adopts, the fixed error-to-code assignment; $T_1$ damping leaks it back, and the dephasing contracts the code block's two $\langle\bar X\rangle$ elements beside the first population cell, the $(0000,1111)$ pair, which no pump transition reaches. (b) Both curves against time $t$ in $\mu$s: $\langle P_{\mathcal{C}}\rangle$, the code-space population, stays above $0.995$ while $\langle\bar X\rangle$ decays with $T_X\approx20\;\mu\mathrm{s}$ to its dotted $1/e$ crossing, $60.8\times$ faster than $T_{\rm gap}$. The inset is the population deficit $1-\langle P_{\mathcal{C}}\rangle$ over the $4\;\mathrm{ms}$ run on a logarithmic axis, its late slope the gap rate.}
\label{fig:gap_vs_logical}
\end{figure}

We prepare $|\psi_0\rangle = (|0000\rangle + |1111\rangle)/\sqrt{2}$, read out by the conjugate logical $\bar X = X_1 X_2 X_3 X_4$, the operator acting on the encoded information. The same separation holds across the plaquette's wiring census, Fig.~\ref{fig:census}: the densest piles sit at the two ends, the fastest at the bare damping time $T_1$ and the slowest three orders above $T_X$. The plaquette is distance one: its stabilizer flags the bit flips the pump returns. Single-qubit dephasing damps $\langle \bar X\rangle$ without altering $\langle P_{\mathcal{C}}\rangle$, Fig.~\ref{fig:gap_vs_logical}(a).

We evolve $|\psi_0\rangle$, Sec.~S6, Fig.~\ref{fig:gap_vs_logical}(b): the code-space population deficit $1-\langle P_{\mathcal C}\rangle$ decays on $T_{\rm gap}=6.1T_1$ against the coherence's $T_X\approx20\;\mu\mathrm{s}$. On the gap's clock this wiring clears break-even; on the coherence's it fails. Against one bare qubit's $0.4$ in $\eta$, $T_{\rm gap}$ exceeds $T_X$ by $60.8\times$, falling to $27\times$ at the per-cycle error budget of Sec.~S10 and $4.5\times$ at demonstrated fidelities. The two rates belong to different modes; earlier work reports a pre-asymptotic discrepancy~\cite{mori2020,mori2023}.

The amplified end rides on the ideal four-body dissipator of Sec.~S10, which no demonstrated circuit element provides. Both extremes of the seven named wirings of Sec.~S7 are non-bijective and so lie outside the census of bijections, whose slowest assignment separates by $2015\times$. Over all $8!$ assignments the slow mode is a code-block population mode, carrying above $99.999\%$ of its weight. The wirings that amplify the gap hold a classical bit rather than a memory: all zeros against all ones, a bit-flip clock the wiring sets against a phase living the bare time. The span grows with the repetition chain, to $1.1\times10^9$ at five qubits, Sec.~S6.

\begin{figure}[!htbp]
\centering
\includegraphics[width=\columnwidth]{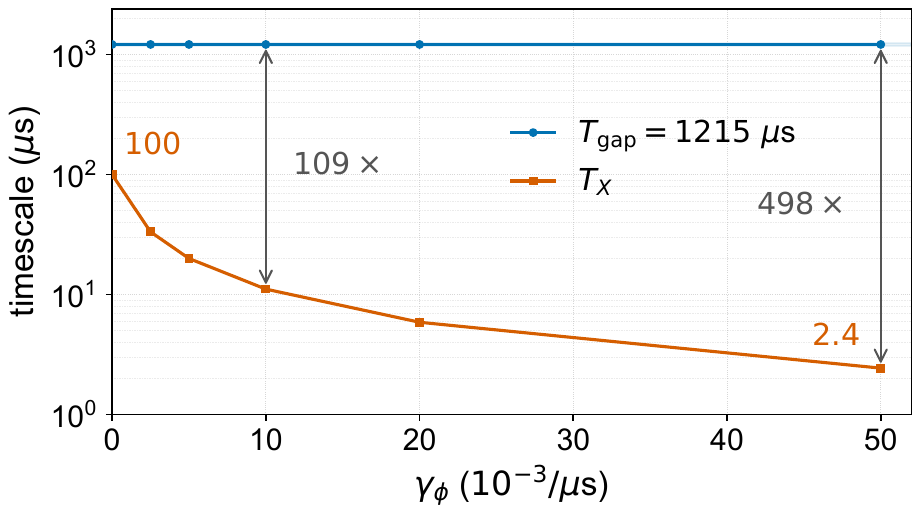}
\caption{The overestimate of the coherence lifetime $T_X$ grows with the commuting noise on the same plaquette, at the pump rate $\kappa = 2\;\mathrm{MHz}$ and $T_1 = 200\;\mu\mathrm{s}$, where the noise is pure $\sigma^z$ dephasing at $\gamma_\phi$. The two timescales against $\gamma_\phi$ in units of $10^{-3}/\mu\mathrm{s}$ on a logarithmic vertical axis: $T_{\rm gap}$, the gap time of the code-space population, is flat at $1215\;\mu\mathrm{s}$, while $T_X$, the logical coherence's decay time, falls from $100\;\mu\mathrm{s}$ at $\gamma_\phi=0$ to $2.4\;\mu\mathrm{s}$, so the separation between them widens from $12\times$ to the $498\times$ of the right-hand arrow; the left-hand arrow marks it at $0.01/\mu\mathrm{s}$, twice the rate of Fig.~\ref{fig:gap_vs_logical}, where the switch scan of Sec.~S7 sits, $109\times$.}
\label{fig:mechanism}
\end{figure}

To localize the cause, we rotate only the noise type at a fixed total rate, Fig.~\ref{fig:mechanism} and Sec.~S7. At twice the rate used above, $0.01/\mu\mathrm{s}$, $\sigma^z$ dephasing holds the two timescales $109\times$ apart, $T_{\rm gap} = 1215\;\mu\mathrm{s}$ against $T_X \approx 11.1\;\mu\mathrm{s}$; $\sigma^x$ bit-flip collapses the separation to $1.4\times$, $T_{\rm gap} = 28.6\;\mu\mathrm{s}$ against $T_X \approx 20.3\;\mu\mathrm{s}$. The collapse is not protection: a bit flip commutes with $\bar X$, so the coherence survives at the $T_1$-limited $100\;\mu\mathrm{s}$ without the pump, which shortens it by damping the code-error coherences, Sec.~\ref{sec:phase}. The gap tracks the syndrome-flipping sector, not where the logical information lives. Over bijections the $\sigma^x$ census spans $1.7\times$, Sec.~S7. The flatness reflects the population closure, App.~\ref{sec:appA}.

\section{The coherence ceiling}
\label{sec:feasibility}
The ceiling's envelope holds at every time and bounds a specified contrast rather than an arbitrary observable. Take $\Delta=(\rho_+-\rho_-)/2$, where $\rho_\pm$ are the opposite-phase superpositions of codewords differing on the $w$ sites of the nonzero bit mask $m$. Its coherences $|x\rangle\langle x\oplus m|$ span the sector $\mathcal S_m$, and $\|\Delta\|_1=1$. For zero Hamiltonian, independent dephasing, amplitude damping, and the basis-state resets of Sec.~\ref{sec:separation}, App.~\ref{sec:appF} proves the envelope at every time: $\|\Delta(t)\|_1\le e^{-\Gamma_w t}\|\Delta(0)\|_1$, with $\Gamma_w=2w\gamma_\phi+w/(2T_1)$. Any normalized phase readout $C_O(t)=\mathrm{Tr}[O\Delta(t)]$, $\|O\|_\infty\le1$, is bounded by this trace distance, so its long-lived component obeys Eq.~\eqref{eq:eta_bound}. Reset rate and wiring do not enter; the drive tolerance of Sec.~III protects the gap, not the bound. The reset loss is diagonal, and the damping clears only common excitations, so $\mathcal S_m$ survives both.

On the plaquette's opposite GHZ inputs, $|0000\rangle\langle1111|$ is an exact generator eigenoperator, giving $T_X=T_2/4=20\;\mu\mathrm{s}$. The envelope is the bare qubit's own line over the weight,
\begin{equation*}
T_X\;\le\;\frac{T_2}{w},\qquad \frac{1}{T_2}=2\gamma_\phi+\frac{1}{2T_1},
\end{equation*}
the lifetime of a bare weight-$w$ contrast under independent local noise~\cite{dur2004,aolita2008} and a steeper bound under correlated noise~\cite{monz2011}. At $w\ge2$ it falls below one bare qubit's coherence, the break-even point. Thermal population divides the bound by $1+2n_{\rm th}$, with $n_{\rm th}$ the mean thermal occupation, Sec.~S2.

A gate-based cycle attains the bound: one parity round followed by a conditional return has Kraus operators $P_+$ and $X_1P_-$. Apart from the independent resets of Sec.~S7, it decays at the unpumped rate under dephasing and at the $T_1$-limited $T_X=2T_1/w$ under bit flip, Sec.~S10. A repetition chain reproduces it to float precision through eight qubits, Sec.~S2.

The hypothesis is a pump and a contrast, not a code class: the two-branch pumps of Sec.~S3 fall outside it. The bound complements the self-correction, memory-lifetime, thermodynamic-limit, and preparation literatures~\cite{kapit2015,kumar2026,dunneweber2026,shtanko2025,bravyi2009,terhal2015,alicki2009,brown2016,dengis2014,konig2014}. A preparation convergence certificate is the gap's remaining role~\cite{verstraete2009,fang2025,temme2010}, Secs.~S4 and S5.

\section{Escape, and its price}
\label{sec:discussion}
The ceiling holds for pumps that cannot carry the phase. One jump returns both error branches at once, so the code block stays intact while the coherence escapes the ceiling. A jump $|c_0\rangle\langle e_0|+|c_1\rangle\langle e_1|$ transfers the error-error coherence to the code coherence with coefficient $\kappa$; every rank-one jump of this class has a vanishing gain term on off-diagonal contrasts. Both are machine-checked in Lean~4, Sec.~S8. A rank-two pump realizing that return sustains the five-qubit code's logical coherence at $T_X=1059.6\;\mu\mathrm{s}$: $\eta=5.30$ on the no-$T_1$ convention of Sec.~S3, $2.92$ once $T_1$ joins the noise, beyond break-even. A coherent return can also saturate the gap, Sec.~S3. Where the pump outruns the dephasing, $T_X\propto\kappa/\gamma_\phi^2$, the law that a continuous implementation already shows~\cite{reiter2017}. Here it follows from the spectral projectors onto the pump's kernel and onto its complement, the states the pump leaves alone.

The structural price is distance. At code distance $d=1$ no single-qubit recovery exists, and only the intrinsic bit-flip invariance at $\eta=2/w$ remains; at $d=2$ errors are detectable but not all correctable, and a detect-only recovery leaves the logical at the unpumped rate under $\sigma^z$ dephasing, $\eta=0.11$, Sec.~S3. The branch-correcting coherent return is first realized at $d=3$, and there it is code-dependent: above the ceiling on the distance-three codes of Sec.~S3, below it yet above its bare-qubit line of $0.222$ on the toric code at lateral size $L=3$, $\eta=0.286\pm0.023$. It survives at distance four, at $0.76 \pm 0.05$ times the distance-three lifetime on resources matched in trajectory count and fit horizon. The ratio remains below unity at $4.8\sigma$ on the fitted error, Sec.~S3.

Fig.~\ref{fig:boundary} marks that price with its green diamonds. Every phase-blind example sits below one bare qubit's line, and on a code detecting every single-qubit error the law $1/T_X=n\gamma_\phi+n/(2T_1)$, with $n$ the qubit count, carries no pump rate. That makes the law general at every distance above one, Sec.~S2. The gap has no protection to fail to measure.

The gate budget binds harder: the escape survives at Sec.~S10's required fidelities, not at the demonstrated ones, and the transmon leakage it leaves open has a hardware precedent, leakage reduction concurrent with measurement~\cite{xin2026}. A matched-budget calculation in Sec.~S3 puts both phase-preserving routes on one correction budget, $\kappa=1/\tau_{\rm cyc}$, the gate cycle's reset rate, and $T_1$ in the noise: measurement-based recovery reaches $\eta=5.59$ against the autonomous pump's $2.92$.

\section{Conclusion}
\label{sec:conclusion}
This memory runs on two independent clocks, the analogs of a qubit's $T_1$ and $T_2$: the wiring moves the population time, the phase keeps the bare $T_2$ divided by the weight. Shadow lattices~\cite{kapit2016}, coherent-return repairs~\cite{rojkov2025}, and measurement-based recovery~\cite{sivak2023,acharya2023} handle the phase the pump discards, so they are assessed on their own readouts, not against this bound.

The result is a two-clock test~\cite{pal2022}: a platform that records the code-space population, the check's parity $(1+\langle\Pp\rangle)/2$, and the logical coherence runs it on data it holds. On the published model of Shirol et al.~\cite{shirol2026}, the slowest relaxation time is $880\;\mu\mathrm{s}$ against a same-model logical coherence of $333\;\mu\mathrm{s}$ and a reported lifetime of $196\;\mu\mathrm{s}$; permuting the returns of three branches moves it between $186$ and $1415\;\mu\mathrm{s}$ while the coherence moves only $184$ to $336\;\mu\mathrm{s}$. Within the rotated-surface family of Sec.~S3 the escape's margin is consistent with $1/d$, and the pump rate needed to clear break-even grows with distance. Every result here is a theorem, an exact computation, or a fit whose scatter is disclosed. Reading a generator's gap as protection needs the second clock beside it.

\begin{acknowledgments}
This work was supported by the National Natural Science Foundation of China under Grant No.~12674609 and by the Guangdong Provincial Key Laboratory of Computility Microelectronics under Grant No.~2024B1212010007. All authors contributed equally to this work. The authors used DeepSeek-V4.1-Flash for text polishing, the Lean derivations, and the debugging of individual numerical values.
\end{acknowledgments}

\section*{Data Availability}
The supplemental derivations, simulation code, machine-checked formalizations, and numerical output files are publicly available on Zenodo at \url{https://doi.org/10.5281/zenodo.22869505}.

\nocite{zanardi1997,lidar1998,higgott2021,gordeychuk2026,nielsen2002,blais2021,sete2015,cohen2014,dennis2002,viola1999,fang2025,george2026doeblin,liu2016,temme2010,lescanne2020,royer2020,denevee2022,reiter2017,parrondo2015,maruyama2009,sagawa2008,sarma2013}
\bibliography{refs}

\appendix

\section{Liouvillian spectrum details}
\label{sec:appA}
The Liouvillian $\mathcal{L}[\rho] = -i[H,\rho] + \sum_k \mathcal{D}[L_k]\rho$~\cite{gorini1976,lindblad1976} is vectorized to a $16^2 \times 16^2$ superoperator on the four-qubit plaquette, at zero Hamiltonian, the protocol of Sec.~S6. The pump acts only on the $P_-$ block, so its zero eigenspace has dimension 64 and the noise lifts this to the unique steady state $|0000\rangle$, the only parity-even string with no excited qubit to decay. The code space is eight-dimensional, three logical qubits at distance one, so the surviving constraint is even parity.

The slowest nonzero mode has rate $1/T_{\rm gap}$ with $T_{\rm gap}=1215\;\mu\mathrm{s}$, a population mode; the next-slowest the $T_1$-limited $2/T_1 = 1/(100\;\mu\mathrm{s})$, $12.2\times$ faster, Sec.~S6; the deposits carry rates and each time here is the inverse of one. The slowest rate is stable across the dephasing scan, and the drive range over which the gap mode and the pump-independence of $T_X$ survive is deposited there.

The flatness is structural. The population sector, the gap-setting sector shown in Fig.~\ref{fig:gap_vs_logical}(a), is closed under the pump, the $T_1$ gain and loss terms and dephasing, so that mode is a population redistribution mediated by $T_1$ leakage and pump return, not a syndrome-flipping relaxation. Its eigenvector carries above $99.999\%$ of its weight on the code-space block in the Euclidean norm, so the gap eigenvalue carries no $\gamma_\phi$ dependence at all; Sec.~S8 machine-checks the dephasing closure, the per-qubit channel fixing every population entry. The mode is the slowest nonzero rate of the classical chain the diagonal populations $p_x=\langle x|\rho|x\rangle$ obey,
\begin{align}
\frac{dp_x}{dt}={}&\frac{1}{T_1}\Bigl[\sum_{i:\,x_i=0}p_{x\oplus e_i}-|x|\,p_x\Bigr]\nonumber\\
&+\kappa\Bigl[\mathbf 1_{\mathcal C}(x)\,p_{\mathcal{W}^{-1}(x)}-\mathbf 1_{\mathcal C^\perp}(x)\,p_x\Bigr],
\label{eq:popchain}
\end{align}
with $e_i$ the unit bit at site $i$, $|x|$ the Hamming weight and $\mathcal{W}$ the wiring of the eight complement basis states onto code states. The Liouvillian's gap matches that rate to below $10^{-13}$ on the cyclic wiring, and only the gain term carries $\mathcal{W}$, so the identification is a property of the reduction rather than of one wiring, Sec.~S8. The gap is a function of the wiring alone: over all $8!$ bijections Eq.~\eqref{eq:popchain}'s $T_{\rm gap}$ spans $200$ to $40300\;\mu\mathrm{s}$, Sec.~S6. The mode is diagonal in the computational basis and carries exactly zero weight on the GHZ contrast $X_1X_2X_3X_4$.

Across the codes characterized here the slow mode is the population relaxation of the logical direction the noise leaves alone, and rotating the noise to $\sigma^x$ exchanges which direction that is, Secs.~S2 and S3.

Every observable decomposes into eigenmodes of $\mathcal{L}$, each decaying asymptotically at a rate at least the gap, so the gap never overstates an asymptotic rate. The population sector is measured to relax on it, Sec.~S6.

The boundary mechanism covers in-code-space population redistribution: channels with a component $E\mathcal{C}\subseteq\mathcal{C}$, such as the excited-state projector $\Pi_i=|1\rangle\langle1|$ inside amplitude damping's no-jump part, contribute to $\mathcal{L}\big|_{\mathcal{C}}$ while leaving the gap-setting sector untouched; under pure $T_1$ decay that contraction sets the coherence decay at $\Gamma_4=w/(2T_1)=2/T_1$ on the plaquette, the $w=4$ case of Eq.~\eqref{eq:eta_bound}, the residual $12\times$ of Fig.~\ref{fig:mechanism} at $\gamma_\phi=0$. Without the pump that separation is the $10\times$ the noise alone sets, stretched by the cyclic wiring to the $60.8\times$ of Secs.~\ref{sec:kapit} and S6: the overestimate is a noise-side separation between population and coherence modes, amplified by the pump's metastability, its slow modes outliving the rest~\cite{macieszczak2016}.

\section{Proof of the pump-invisibility statement and the contrast envelope}
\label{sec:appF}
Let $P_\pm = (I \pm g)/2$ project onto the $g = \pm 1$ eigenspaces, with $\mathcal{C} = \mathrm{ran}\,P_+$. The Diehl--Kraus pump consists of jumps $L_\alpha = \sqrt{\kappa}\,|\psi_\alpha\rangle\langle\phi_\alpha|$ with $|\phi_\alpha\rangle \in \mathrm{ran}\,P_-$ and $|\psi_\alpha\rangle \in \mathcal{C}$, so $L_\alpha$ annihilates the code and maps $\mathrm{ran}\,P_-$ into $\mathcal{C}$. Each of the dissipator's two terms then carries a factor $L_\alpha P_+$ or its adjoint,
\begin{equation}
L_\alpha P_+ = 0,\quad L_\alpha P_+\rho P_+ L_\alpha^\dagger = 0,\quad P_+L_\alpha^\dagger L_\alpha P_+ = 0,
\label{eq:pumpboundary}
\end{equation}
so $\mathcal{D}[L_\alpha]\rho = L_\alpha \rho L_\alpha^\dagger - \tfrac12\{L_\alpha^\dagger L_\alpha, \rho\}$ annihilates every code-supported state, $\mathcal{D}[L_\alpha](P_+\rho P_+) = 0$. This is the boundary identity of Sec.~\ref{sec:separation}.

On the code-space block Eq.~\eqref{eq:pumpboundary} leaves such a channel acting as the noise alone. A component commuting with $g$ commutes with $P_\pm$, hence maps the block into itself, $P_+EP_+ = EP_+$. Writing $F_k = P_+E_kP_+$ for the block-restricted component, its adjoint action on block operators and its normalization are
\begin{equation}
\rho \;\longmapsto\; \sum_k F_k^\dagger\,\rho\,F_k,
\qquad \sum_k F_k^\dagger F_k = P_+,
\label{eq:blockchannel}
\end{equation}
the second identity the completeness relation restricted to the block: one component's outflow is cancelled by the rest, so the block's generator is the noise's own, Eq.~\eqref{eq:blockchannel}.

A noise-only adjoint identity applies to $X/Y/I$ tensors. Additional $Z$ factors require coupled dynamics: amplitude damping gives $\mathcal D^\dagger[\sigma^-/\sqrt{T_1}](Z)=(I-Z)/T_1$, and the four-qubit amplitude-damping code, whose $Z$-type check $Z_1Z_2Z_3Z_4$ flips under the damping, evades this half of the bound~\cite{leung1997}. For such codes the envelope argument below, not this adjoint identity, carries the $T_1$ half. Nor does the weight of one Pauli establish the sector hypothesis for an arbitrary encoded state, gauge initialization or frame-changing recovery. At the generator level the hypothesis is that every noise jump commutes with the code projector, $L_aP_{\mathcal C}=P_{\mathcal C}L_a$. The Hamiltonian preserves the code and its complement, which with $L_\alpha P_{\mathcal C}=0$ for the pump makes the evolution of any initially code-supported state the noise and code-Hamiltonian evolution within the code space, independently of $\kappa$; this is sufficient for the Remark, whereas invariance under a jump alone does not ensure invariance under its dissipator. The population-sector analysis of App.~\ref{sec:appA} identifies the plaquette gap, and for a code-error coherence $|e\rangle\langle c|$ the pump gain vanishes because $\langle c|\phi_\alpha\rangle=0$.

Under the pump alone every code-supported density matrix is stationary, a code space of dimension above one carrying several, Spohn's commutant criterion~\cite{spohn1977}; the noise lifts the degeneracy, and App.~\ref{sec:appA} resolves which mode sets the gap.

For the lifetime theorem, specialize to basis-state jumps $L_\alpha=\sqrt{\kappa_\alpha}|\psi_\alpha\rangle\langle e_\alpha|$, normalized targets and nonnegative rates. Put $r_x=\sum_{\alpha:e_\alpha=x}\kappa_\alpha$. For $B_{xy}=|x\rangle\langle y|$, $x\ne y$, direct application of the full generator yields
\begin{align}
\mathcal L(B_{xy})={}&-\left[2\gamma_\phi w+\frac{|x|+|y|}{2T_1}+\frac{r_x+r_y}{2}\right]B_{xy}\nonumber\\
&+\frac{1}{T_1}\sum_{i:x_i=y_i=1}B_{x\setminus i,y\setminus i},
\label{eq:sector_generator}
\end{align}
where $m=x\oplus y$, $w=|m|$, and $x\setminus i$ clears bit $i$. The reset gain vanishes on every off-diagonal $B_{xy}$ because a computational input cannot match both strings, while a jump reaching both returns the contrast as code-block coherence at the reset rate. On $|e_0\rangle\langle e_1|$ with $e_0\neq e_1$, the single-branch jump $L=\sqrt\kappa\,|c\rangle\langle e|$ and the two-branch $L=\sqrt\kappa\bigl(|c_0\rangle\langle e_0|+|c_1\rangle\langle e_1|\bigr)$ give
\begin{align}
L\,|e_0\rangle\langle e_1|L^\dagger&=0,
\label{eq:gainvanish}\\
L\,|e_0\rangle\langle e_1|L^\dagger&=\kappa\,|c_0\rangle\langle c_1|.
\label{eq:gainreturn}
\end{align}
Eq.~\eqref{eq:gainvanish} is the vanishing gain of Eq.~\eqref{eq:sector_generator}; Eq.~\eqref{eq:gainreturn} is the coherent return of Sec.~\ref{sec:discussion}. Damping clears only common excited bits, preserving $m$; the reset loss is diagonal. Thus $\mathcal S_m$ is invariant even for superposition targets.

Write $A=\sum_xa_x|x\rangle\langle x\oplus m|=\mathrm{diag}(a)X_m$, where $X_m$ is a permutation matrix, so $\|A\|_1=\sum_x|a_x|$. The coefficient generator has nonnegative off-diagonal entries and column sum
\begin{align}
-\Gamma_w-\frac{r_x+r_{x\oplus m}}{2}&\le-\Gamma_w,
\qquad \Gamma_w=2w\gamma_\phi+\frac{w}{2T_1},
\label{eq:colsum}\\
\|\Delta(t)\|_1&\le e^{-\Gamma_w t}\|\Delta(0)\|_1,\qquad t\ge0,
\label{eq:envelope}
\end{align}
so by Eq.~\eqref{eq:colsum} the upper right derivative of the coefficient norm is at most $-\Gamma_w\sum_x|a_x|$, and integrating gives the envelope of Eq.~\eqref{eq:envelope}, $T_X\le1/\Gamma_w$ for the slowest component,
\begin{equation}
\eta\equiv\frac{T_X}{T_1}\le\frac{1}{T_1\Gamma_w}=\frac{2}{w+4wT_1\gamma_\phi}=\frac{T_2}{wT_1}\le\frac{2}{w},
\end{equation}
which with $1/T_2 = 2\gamma_\phi+1/(2T_1)$ is the inequality chain of Eq.~\eqref{eq:eta_bound}, the middle expression the $w$-fold contrast of that equation, its $T_1$ half the per-factor contraction above and its dephasing half the $2w\gamma_\phi$ of $\Gamma_w$. That half carries a factor two because the channel is the Kraus pair $\sqrt{1-\gamma_\phi}\,I$ and $\sqrt{\gamma_\phi}\,Z$, which multiplies an off-diagonal element by $1-2\gamma_\phi$ per step: the per-weight rate is one qubit's total decoherence rate, $1/T_2=1/T_\phi+1/(2T_1)$ with $T_\phi=1/(2\gamma_\phi)$, so at the headline parameters $T_2=80\;\mu\mathrm{s}$ and the plaquette's $T_X=T_2/4=20\;\mu\mathrm{s}$. Bare and weighted, the relation is one: a physical qubit ties its two clocks by $1/T_2=2\gamma_\phi+1/(2T_1)$, a weight-$w$ contrast by the same relation times the weight, $\Gamma_w=w/T_2$, so the ceiling of Eq.~\eqref{eq:eta_bound} is the bare line $\eta_{\rm bare}=T_2/T_1$ over the weight, exact at $w=1$ and tighter at $w\ge2$. This controls transients as well as asymptotic exponents, without asserting a single exponential for every readout, with every step shown in Sec.~S2 of the Supplement.

For a codeword pair with no common excited site and no reset input, $r_x=r_y=0$ and $|x|+|y|=w$, so Eq.~\eqref{eq:sector_generator} leaves $B_{xy}$ an exact eigenoperator at one bare-qubit rate spread over $w$,
\begin{equation}
\mathcal{L}(B_{xy})=-\frac{w}{T_2}\,B_{xy},
\qquad \frac{1}{T_2}=2\gamma_\phi+\frac{1}{2T_1},
\label{eq:exacteig}
\end{equation}
Eq.~\eqref{eq:exacteig} holds for the plaquette's own $x=0000$, $y=1111$ among them, with $T_X=T_2/4=20\;\mu\mathrm{s}$ exact rather than fitted.

For repeated recovery, a finite noise interval preserves $\mathcal S_m$, every column of its coefficient matrix summing to at most $e^{-\Gamma_w\tau}$, one interval of the envelope, so $n$ intervals contract the trace norm by at most $e^{-n\Gamma_w\tau}$.

A complete $Z$-syndrome measurement followed by Pauli corrections has Kraus operators $U_sP_s$: diagonal $P_s$ retains or removes matrix elements and $U_s$ flips the matching bits in both indices, preserving their XOR. The resulting recovery, a completely positive trace-preserving map, preserves $\mathcal S_m$ and contracts the trace norm, and iterating noise followed by recovery proves the envelope for any number and durations of these cycles, a final decoder included. Non-diagonal coherent control, superposition-valued reset inputs and unaccounted external memory fall outside this proof.

The bare-noise adjoint tensor identities remain exact, Sec.~S8: for $X/Y/I$ factors, with $\Phi_{T_1}$ the finite-interval damping channel, $\Phi_{T_1}^\dagger(\bar X)=(1-p)^{w/2}\bar X$, while $Z$ obeys $\Phi_{T_1}^\dagger(Z)=(1-p)Z+pI$, and neither alone proves the pumped or recovered dynamics.

The pump boundary and noise-only algebra are machine-checked in Lean~4. Of the sector argument above, the coefficient-level chain, the trace-norm identity, the column sum and the envelope are machine-checked for an arbitrary number of qubits, the population block's internal ordering being the one unformalized step, Sec.~S8; its composition across noise intervals and recovery cycles is analytic, with independent finite checks in Sec.~S6. The conditional iteration lemmas of Sec.~S8 carry hypotheses and do not certify arbitrary recovery.

\end{document}